\documentclass[twocolumn, amssymb, amsmath, aps, prl, showpacs, 10pt]{revtex4-1}
\usepackage[dvips]{graphicx}
\usepackage[dvips]{color}

\usepackage{xcolor}
\usepackage{hyperref}
\hypersetup{
    colorlinks=true,
    linkcolor=blue,
}

\begin{document}
\title{Robustness of helical magnetic structure under external pressure in Fe-doped MnNiGe alloy}
\author{S. C. Das$^1$}
\author{J. Sannigrahi$^{2,3}$}
\author{K. Mandal$^1$}
\author{P. Dutta$^4$}
\author{S. Pramanick$^1$}
\author{D. Khalyavin$^2$}
\author{D. T. Adroja$^{2,5}$}
\author{S. Chatterjee$^1$}
\email{souvik@alpha.iuc.res.in}
\email{souvikchat@gmail.com}
\affiliation{$^1$UGC-DAE Consortium for Scientific Research, Kolkata Centre, Sector III, LB-8, Salt Lake, Kolkata 700 106, India}
\affiliation{$^2$ISIS Facility, Rutherford Appleton Laboratory, Chilton, Didcot OX11 0QX, United Kingdom}
\affiliation{$^3$Department of Physics, Loughborough University, Loughborough LE11 3TU, United Kingdom}
\affiliation{$^4$New Chemistry Unit and School of Advanced Materials, Jawaharlal Nehru Centre for Advanced Scientific Research, Jakkur, Bangalore 560064, India}
\affiliation{$^5$Highly Correlated Matter Research Group, Physics Department, University of Johannesburg, PO Box 524, Auckland Park 2006, South Africa}
\begin{abstract}
We have investigated the detailed magnetic structure of a Fe-doped MnNiGe alloy of nominal composition MnNi$_{0.75}$Fe$_{0.25}$Ge by ambient and high pressure (6 kbar) neutron powder diffraction study. The alloy undergoes a martensitic phase transition between 230 K and 275 K. The low-temperature martensite phase orders antiferromagnetically with helical modulation of Mn-spins. At 1.5 K, the incommensurate propagation vector $\bf k$ is found to be (0.1790(1),0,0), and it remains almost unchanged with temperature under ambient condition. The Application of external pressure ($P$) reduces the martensitic transition temperature and results in a significant change in the incommensurate magnetic propagation vector. At 10 K, with 6 kbar of external pressure, the incommensurate magnetic structure of the alloy stabilizes with $\bf k$ = (0.1569(1),0,0). Interestingly, the strong temperature dependence of $\bf k$ in the presence of external $P$ has also been observed. 

\end{abstract}
\maketitle
The magnetic equiatomic alloys (MEAs) with general formula MM$^{\prime}$X (where M and M$^{\prime}$ are transition metals, and X is a nonmagnetic $sp$ element) are in the forefront of active research due to the observation of various magnetofunctional properties~\cite{zhang-apl,Koyama-fsma,zhang-jpd,pd-epl,ali-apl,Trung-apl1,pd-jmmm,liu-nc,Caron-prb1,Dincer-jalcom1,Trung-mncoge,Wang-mncoge,pd-prb1,pd-jpd2,km-jpd1,km-jap1,km-jalcom1,scd-jpd1,scd-jmmm1}. MnNiGe is among the most studied alloys of MM$^{\prime}$X family and crystallizes in a Ni$_2$In-type hexagonal structure (space group $P6_3/mmc$, with transition metals at 2$a$(0,0,0) and 2$d$($\frac{1}{3},\frac{2}{3},\frac{3}{4}$) sites, whereas, Ge atoms at 2$c$($\frac{1}{3},\frac{2}{3},\frac{1}{4}$) site) and undergoes a martensitic phase transition (MPT) around 470 K to the TiNiSi-type orthorhombic structure (space group $Pnma$, with all atoms at 4$c$($x$,$\frac{1}{4}$,$z$) position)~\cite{zhang-apl,zhang-jpd,pd-epl,liu-nc}. Further cooling results in a magnetic transition (paramagnetic to antiferromagnetic) in the alloy, and it becomes antiferromagnetically ordered below 346 K~\cite{zhang-apl,zhang-jpd,pd-epl,liu-nc}. The previous neutron diffraction studies confirmed the spiral nature of the antiferromagnetic order (with an ordered moment in the Mn site only) at ambient condition~\cite{bazela-pssa1,bazela-pssa2,bazela-pt,Penca-pt}. Among others, Fe-doping in Mn or Ni-site is particularly crucial due to the broad tunability of MPT~\cite{liu-nc,pd-epl,pd-prb1,pd-jpd2,km-jpd1}.  Our recent work reveal several exciting properties of Fe-doped MnNiGe alloys~\cite{pd-epl,pd-prb1,pd-jpd2,km-jpd1}. Though various doping studies have been performed to explore the functional features at the ambient pressure of these materials, there are very few efforts to address their magnetic structure~\cite{bazela-pssa1,bazela-pssa2,bazela-pt,Penca-pt}. As magnetic structure plays a crucial role behind the observation of different magneto-functional behaviors, it is, therefore, pertinent to investigate the exact magnetic structure of the MEAs. 

\par
In the present work, we aim to focus on the magnetic structure of the Fe-doped MnNiGe alloy of nominal composition MnNi$_{0.75}$Fe$_{0.25}$Ge in ambient as well as in high-pressure condition through detailed neutron powder diffraction studies. To the best of our knowledge, this is the first-ever attempt to probe the magnetic structure of MEA in the presence of external pressure. Our result reveals a helical incommensurate antiferromagnetic structure of this alloy below the MPT and a significant decrease in incommensurate propagation vector has been observed in the presence of external pressure.

\par
The Fe-doped MnNiGe alloy of nominal composition MnNi$_{0.75}$Fe$_{0.25}$Ge was prepared in polycrystalline form by argon arc melting method~\cite{km-jpd1}. The neutron powder diffraction (NPD) measurements of the well characterized alloy were performed in the cold neutron (time of flight) diffractometer WISH at ISIS Facility (Rutherford Appleton Laboratory) in the UK~\cite{Chapon-nn}. For ambient pressure experiment, the powder sample was mounted in a 6 mm vanadium cell and cooled down to 1.5 K using a standard He-cryostat. We used TAV6 gas cell (Ti-6Al-4V-alloy) with the internal diameter 7mm, attached to the automated intensifier system for high-pressure NPD experiments, where high-pressure Helium gas was used as pressure media. In this method, one can apply up to 10 kbar uniform pressure ($P$) on the sample. The pressure cell was inserted into the He-cryostat and cooled down to 10 K. Data were collected at different constant temperatures ($T$) in the warming cycle. FULLPROF software package was used for refining the NPD data~\cite{jr-pb}.

\begin{figure}[t]
\centering
\includegraphics[width = 8 cm]{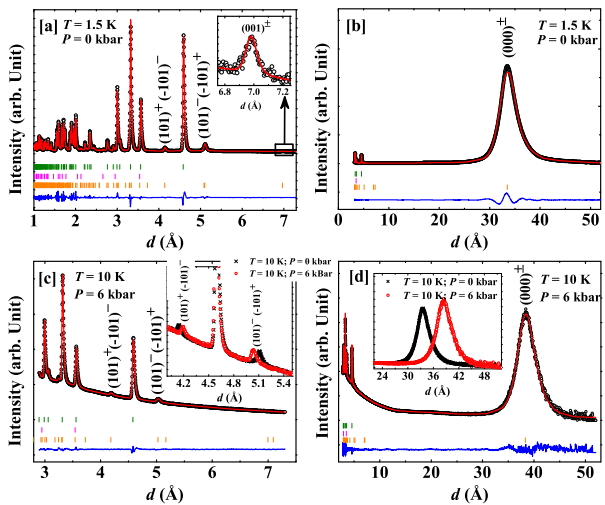}
\caption{NPD data recorded at 1.5 K in ambient conditions are plotted in the main panels of (a) and (b) for two different ranges of lattice spacing ($d$) from the detector banks 3-8 and 1-10 of WISH respectively. Inset of (a) indicates the enlarged view of the (001)$^{\pm}$ magnetic peak. Main panels of (c) and (d) depict the high-pressure NPD data in the presence of 6 kbar of external pressure at 10 K for two different ranges of $d$ from the detector banks 3-8 and 1-10 of WISH respectively. Insets of (c) \& (d) show the enlarged view of the magnetic peaks recorded in ambient and high-pressure conditions at 10 K. Here, black circle, red lines, and blue lines indicate the experimental, calculated and difference patterns respectively. The first (olive), second (magenta) and third (orange) vertical bars indicate the peak positions for the orthorhombic nuclear phase, hexagonal nuclear phase, and orthorhombic magnetic phase, respectively.}
\label{npd}
\end{figure}

\begin{figure}[t]
\centering
\includegraphics[width = 8.5 cm]{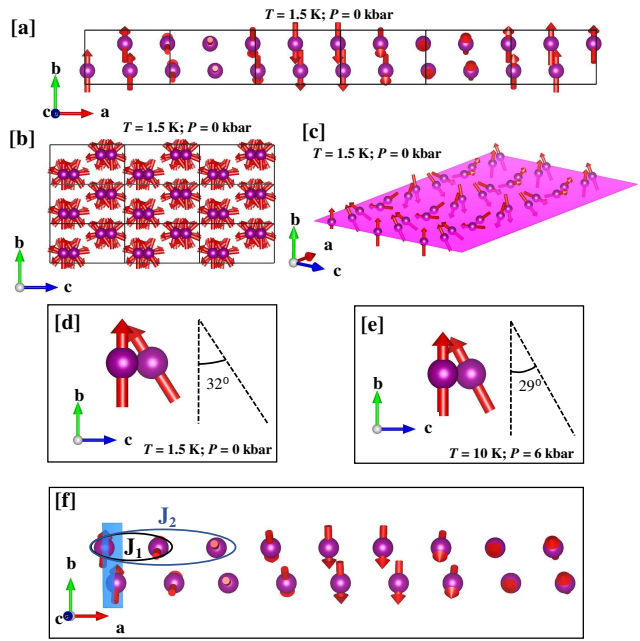}
\caption{(a), (b) and (c) show the illustration of the helical magnetic structure of MnNi$_{0.75}$Fe$_{0.25}$Ge. (d) and (e) depicts the angle between two adjacent magnetic spins in ambient and high-pressure conditions, respectively. For clarity, only the magnetic atoms, Mn, are shown here. (f) shows the exchange interactions present between $bc$ planes along $a$ axis.}
\label{str}
\end{figure}

\par
Isothermal NPD data recorded at different constant $T$ indicate a clear change in nuclear structure from orthorhombic ($Pnma$) to hexagonal ($P6_3/mmc$) while warming. Refinement of 305 K NPD data shows a simple hexagonal crystal structure (see fig.S1 of the supplementary section for details), whereas, an orthorhombic structure of the alloy is evident from the 1.5 K data (see fig.~\ref{npd}(a)). The nuclear structures of both the phases are illustrated in fig.S2 of the supplementary section. In addition to the nuclear reflections, some magnetic satellite reflections have also been observed in the low-$T$ orthorhombic phase. Fig.~\ref{npd} (a) \& (b) depict the NPD pattern of the studied alloy at 1.5 K in ambient conditions for different lattice plane spacing ($d$) ranges. The position of the most intense magnetic reflection is found to be very high ($d \sim$ 33.4 \AA), which is due to a larger value of the magnetic form factor of Mn at larger $d$ or smaller $Q$ ($= \frac{2\pi}{d} \sim$ 0.188 \AA$^{-1}$). All the magnetic reflections observed below the magneto-structural transition temperature can be indexed by the incommensurate propagation vector $\bf k =$ ($k_a$,0,0) with $k_a =$ 0.1790(1) at 1.5 K. We successfully refined the diffraction data at 1.5 K, using the orthorhombic space-group for the nuclear scattering and the non-collinear spin structure for the magnetic reflections. Our analysis also indicates the presence of a high-$T$ hexagonal phase ($\sim$0.48 weight \%) even at the lowest-$T$ of measurements (see table S1 of the supplementary section for refined parameters). The refinement of the magnetic structure was assisted by symmetry arguments based on the space group representation theory~\cite{Bertaut,Izyumov,Izyumov-jmmm}. We tested all possible models for the magnetic structure and confirmed that all Mn atoms are fully ordered, having equal moments at all Mn sites. The wave vector group splits the 4$c$ Mn position into two orbits Mn$_{11}$ (0.0300,0.2500,0.1805), Mn$_{12}$ (0.5300,0.2500,0.3195) and Mn$_{21}$ (0.4700, 0.7500, 0.6805), Mn$_{22}$ (0.9700, -0.2500, -0.1805). The $x$-coordinate of the two Mn sites in each orbit differs by 0.5, implying the phase difference $\frac{k_a}{2}$. The magnetic phase between the orbits is not fixed by the propagation vector and was determined from the refinement. It was possible since the intensity of the $(001)^{\pm}$ satellite was found to be very sensitive to the dephasing between the orbits. The model which provided the best fitting quality (the reliability factor $R_{\rm mag} =$ 4.2\%) was the helical structure with the spins rotating within the $bc$-plane as depicted in fig.~\ref{str} (a), (b), and (c). The parameters of the magnetic structure refined at $T =$ 1.5 K are summarized in Table~\ref{table}.

\begin{table*}[t]
\centering
\caption{Fourier coefficients of the helicoidal magnetic structure of MnNi$_{0.75}$Fe$_{0.25}$Ge, refined at $T =$ 1.5 K for ambient pressure (propagation vector $\bf k =$ (0.1790(1),0,0)) and at $T =$ 10 K for $P =$ 6 kbar (propagation vector $\bf k =$ (0.1562(1),0,0)). The magnetic structure was defined as $(R{\bf v}+iI{\bf w})e^{-2\pi i({\bf k.t}+ \varphi)}$, where $R$ and $I$ are real and imaginary parts of the Fourier coefficients, respectively, $\bf t$ is lattice translation, and $\varphi$ is the magnetic phase. $\bf v$ and $\bf w$ are unit vectors defining the plane of the spin rotation ($bc$-plane in the present case).}
\begin{tabular}{c|c|c|c|c}
\hline
\hline
Mn-site & \multicolumn{2}{c}{Moment size, $R = I$ ($\mu_B$)} & \multicolumn{2}{c}{Magnetic phase, $\varphi$} \\
\hline
&$P$ = 0, $T$ = 1.5 K & $P$ = 6 kbar, $T$ = 10 K & $P$ = 0, $T$ = 1.5 K & $P$ = 6 kbar, $T$ = 10 K\\
\hline
Mn$_{11}$(0.0300,0.2500,0.1805) & 3.09(3) & 3.02(2) & 0.000(0) & 0.000(0)\\
Mn$_{12}$(0.5300,0.2500,0.3195) & 3.09(3) & 3.02(2) & 0.089(5) & 0.078(1)\\
Mn$_{21}$(0.4700,0.7500,0.6805) & 3.09(3) & 3.02(2) & 0.099(1) & 0.088(1) \\
Mn$_{22}$(0.9700,-0.2500,-0.1805) & 3.09(3) & 3.02(2) & 0.188(1) & 0.166(1)\\
\hline
\hline
\end{tabular}
\label{table}
\end{table*}

\par
The model implies that the angle between two adjacent spins of the Mn atoms along the propagation vector is $\alpha \approx$ 32$^{\circ}$ at 1.5 K (see fig.~\ref{str} (d)). Notably, no ordered moment on the Fe atoms has been detected from the NPD analysis. A similar helical magnetic structure has also been reported for a specific $T$ range of Si, Ti, and Cr-doped MnNiGe alloys~\cite{bazela-pssa1,bazela-pssa2,bazela-pt,Penca-pt}. The position of the magnetic reflections remains almost the same with increasing temperature (see fig.S3 (a) and (b) of the supplementary section for details). However, the intensity of the magnetic reflections shows sluggish decrease up to 230 K followed by a sharp drop with further increasing $T$, when the system starts transforming to the hexagonal Ni$_2$In-type structural phase (fig.~\ref{para} (a) and (b)). The magnetic transition temperature of the orthorhombic phase falls within the same range of the structural transition temperature, ({\it i.e.} between 230-275 K) but cannot be reliably determined from the neutron diffraction data since the fraction of this phase rapidly decreases above 230 K. Phase co-existence around the first-order phase transition is a common phenomenon, and the system remains in the metastable state between the two endpoints of the transition, $T^{*}$ and $T^{**}$ (austenite start and austenite finish temperature respectively for the present case)~\cite{Chaikin}. Such phase co-existence gives rise to a superheated low-$T$ martensitic phase (up to $T^{**}$) and plays a pivotal role behind the observation of such a broad range of magnetic and structural transition temperatures. As mentioned above, no change of the magnetic structure was detected, and we successfully refined the neutron diffraction pattern in the entire temperature range with the same helicoidal model. A monotonic increase of the lattice parameters of both the phases has been observed (see fig. S4 in the supplementary section). Not much variation in the propagation vector has been observed with increasing sample $T$ (see inset of fig.~\ref{para} (b)). This is again unlike the Si and Ti-doped MnNiGe alloys, where a significant increase in propagation vector has been reported~\cite{bazela-pssa1,bazela-pssa2,bazela-pt}. However, the behavior is similar to that of the 11\% Cr-doped MnNiGe alloy~\cite{Penca-pt}. We have also estimated $\alpha$ at different temperatures, and only $1^{\circ}$ decrease in its value has been noticed with $T$ changing from 1.5 K to 250 K ($\alpha = 31^{\circ}$ at 250 K). We have checked the existence of any magnetically ordered hexagonal phase. But the low-temperature hexagonal phase fraction is too small to detect any magnetic reflection. In the high-temperature region, especially above the martensitic transition temperature, the absence of any magnetic satellite reflection is clear from our NPD patterns. Such observation confirms that the present alloy transforms from an antiferromagnetically ordered orthorhombic phase to a paramagnetic hexagonal phase. This description at ambient pressure is used as a reference to analyze the effect of pressure on the magnetic order.

\par
In the presence of external $P$, no drastic change has been observed at the high-temperature paramagnetic phase, and hence the high-$P$ nuclear structure bears a strong resemblance to that of ambient pressure. Minute inspection of the high-$P$ NPD data recorded at 10 K confirms the absence of any new set of magnetic Bragg reflection (see main panels of fig.~\ref{npd} (c) \& (d)). Interestingly, the positions of magnetic reflections change significantly as compared to ambient pressure pattern which implies that the new propagation vector is still ${\bf k} = (k_a, 0, 0)$ but with a different value of $k_a$ (see insets of fig.~\ref{npd} (c) \& (d) for the closure look of magnetic reflections for ambient and high-pressure data). Besides, a shift in nuclear reflections also confirms the decrease in both orthorhombic lattice parameters in the presence of an external $P$. At 10 K we can index all magnetic reflections with $k_a =$ 0.1562(1). A non-collinear magnetic structure with helical modulation of Mn spins (similar to ambient pressure magnetic structure) is therefore proposed. We obtain an excellent agreement between observed and calculated intensities with $R_{\rm mag} =$ 8.57 \%. The refined parameters of the magnetic structure at 10 K in the presence of 6 kbar of $P$ are summarized in Table~\ref{table}. The helical magnetic structure constructed from the refinement analysis indicates that the angle between two adjacent Mn-spins reduces to 29$^{\circ}$ at 10 K (see fig.~\ref{str} (e)); {\it i.e.} the alloy is going towards the parallel arrangement of magnetic spins ({\it i.e.} ferromagnetically ordered state) in the presence of external $P$. The restricted views of the NPD patterns at different constant $T$ indicate a significant shift in the position of the magnetic reflections with $T$ (see fig. S3 (c) \& (d) of supplementary material). Besides, a monotonic decrease in the intensities of the magnetic reflections has also been observed with increasing $T$ and eventually vanishes above 200 K. We have successfully refined all NPD data recorded in high-$P$ condition. The shift of magnetic reflections in the presence of external $P$ results in a strong $T$ dependence of propagation vector (around 10 K $\le T \le$ 200 K, see inset of fig.~\ref{para} (b)). $k_a$ decreases monotonically with increasing temperature, and it is found to be 0.1325(8) at 200 K. Unlike the ambient pressure case, about 5$^{\circ}$ decrease in $\alpha$ has been observed for $T$ changing from 10 K to 200 K with $\alpha$ = 24$^{\circ}$ at 200 K. A notable amount of decrease ($\sim$ 60 K decrease) in magneto-structural transition temperature under external $P$ has also been observed (see fig.~\ref{para}(a) phase fraction {\it vs.} $T$ curve in the presence of 6 kbar of $P$). Interestingly, not much change in the Mn-site moment size due to external $P$ has been noticed in the low-$T$ martensitic phase (see Table~\ref{table}). We have also calculated and plotted the integrated intensity of the (000)$^{\pm}$ magnetic reflection as a function of temperature, and it further confirms the clear shift of structural transition temperature in the presence of external pressure and shows one-to-one correspondence with the orthorhombic phase fraction (see fig.~\ref{para}(b)). We also noticed similar temperature variation of the integrated intensities for other magnetic satellite reflections.
\begin{figure}[t]
\centering
\includegraphics[width = 7.5 cm]{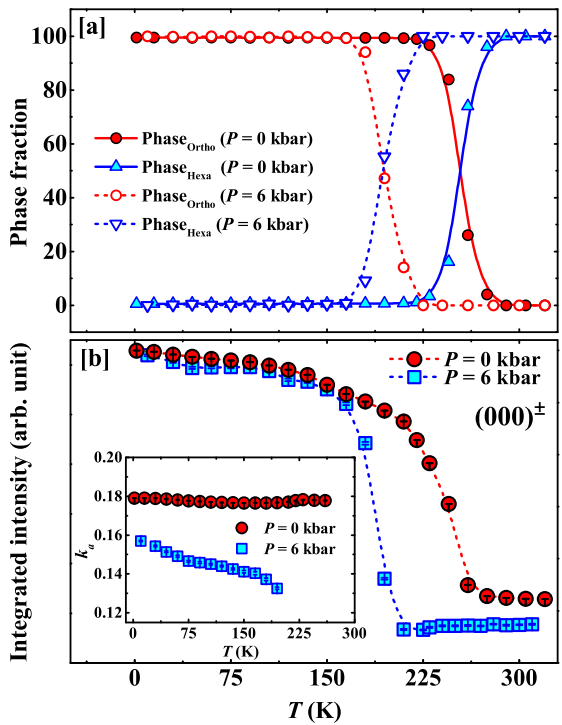}
\caption{Temperature variations of  (a) phase fraction, (b) integrated intensities of $(000)^{\pm}$ magnetic peak in ambient as well as in the presence of 6 kbar of external pressure. Inset of (b) depicts the temperature variation of $k_a$ both under ambient and high-pressure situation.}
\label{para}
\end{figure}

\par
Recent observation of helical magnetic structures in Ti and Cr-doped MEAs have been analyzed based on the available theoretical models~\cite{bazela-pt,Penca-pt,Herpin,Enz}. Competing exchange interactions often lead to such a type of helical magnetic structure. The moment direction rotates by a certain angle while going from one layer to another layer along the direction of the propagation vector. For the presently studied alloy, this rotation of the adjacent spins of Mn-atoms depends strongly on $T$ under high-$P$ condition, while a negligible change has been observed in ambient conditions. The theoretical model proposes the axial next nearest-neighbor Ising model that incorporates anisotropic competing exchange interactions in such a way that the spins are coupled by nearest-neighbor ferromagnetic interaction in a plane perpendicular to the modulation axis of the helical magnetic structure. Along the modulation axis, the spins are coupled by nearest-neighbor ferromagnetic and next-nearest neighbor antiferromagnetic interactions, which together with leads to the helical phase with the modulation vector perpendicular to the ferromagnetic plane. A stable helical structure is obtained for the condition $\cos \alpha = -\frac{J_1}{4J_2}$, where $\alpha$ is the angle between two adjacent spins of Mn atoms and $J_1$, $J_2$ signify the ferromagnetic and antiferromagnetic interactions respectively along the modulation axis (see fig.~\ref{str} (f))~\cite{Enz}. A similar modulated helimagnetic phase has also been observed in Tb, Dy, and Ho rare earth elements where alike competing interactions have also been observed~\cite{tapan}. In the present work, the values of $\alpha$ at different constant $T$ have been estimated both in ambient as well as in high-$P$ conditions, which show good agreement with the relation $\alpha$ = 180$^{\circ}$.$k_a$, used by Penca {\it et al.}~\cite{bazela-pt,Penca-pt}. From the stability criteria of the helical modulation, it is evident that any decrease in $\alpha$ is due to the increase in $J_1/J_2$ ratio. Using the values of $\alpha$ obtained from the NPD analysis, we have calculated the values of $J_1/J_2$ ratio both in ambient as well as in high-pressure conditions and observed a significant increase in $J_1/J_2$ ratio ($\frac{J_1}{J_2} \approx$ 3.38 at 1.5 K under ambient pressure and $\frac{J_1}{J_2} \approx$ 3.52 at 10 K in the presence of $P$ = 6 kbar). This indicates the weakening of antiferromagnetic interaction under external $P$. The observed effect of external $P$ on the magnetic structure is found to be similar to that of the Cr-doping, where a clear decrease in $k_a$ has been observed with increasing Cr-concentration. We also noticed a large decrease in $\alpha$ with increasing sample $T$ under high-$P$ condition. Such a decrease in $\alpha$ implies further increase in $\frac{J_1}{J_2}$ ratio and hence a decrease in antiferromagnetic interaction with increasing sample $T$.

\par
In summary, the present work based on the NPD analysis in ambient as well as in the presence of external $P$ reveals a clear picture of the magnetic structure of the MnNi$_{0.75}$Fe$_{0.25}$Ge alloy. Both ambient and high-pressure data confirm that the low-$T$ orthorhombic phase of the present alloy shows incommensurate antiferromagnetic ordering with a helical spin structure. The axis of the helix is found to be along the $a$-axis throughout the magnetically ordered region. On the other hand, no signature of the magnetically ordered hexagonal phase has been observed. The magnetic spin arrangement of the Mn atoms remains helical even in the presence of external $P$ with no new set of magnetic reflection in the NPD pattern. But, a significant decrease in the propagation vector in the presence of external $P$ has been observed. Such a reduction in the propagation vector plays a crucial role in reducing the angle between the two adjacent Mn spins of the helix and hence to decrease the next-nearest neighbor antiferromagnetic interaction along the direction of the propagation vector. This indicates that external $P$ prefer parallel spin arrangement, and with sufficiently high $P$, ferromagnetic ordering can be achieved.

\par
SC, SCD, and KM would like to thank UK Newton funding and DST-India (SR/NM/Z-07/2015) for the access to the experimental facility and financial support to carry out the experiments (DOI: \href{https://data.isis.stfc.ac.uk/doi/STUDY/103197536/}{10.5286/ISIS.E.RB1820028} and \href{https://data.isis.stfc.ac.uk/doi/STUDY/105606470/}{10.5286/ISIS.E.RB1968022}), and Jawaharlal Nehru Centre for Advanced Scientific Research (JNCASR) for managing the DST-India project. SCD (IF160587) would like to thank DST-India for his inspire fellowship. JS would like to thank the European Unions Horizon 2020 research and innovation program under the Marie Skodowska-Curie grant agreement (GA) No 665593 awarded to the Science and Technology Facilities Council.


%

\newpage

\bf{Supplementary Material for ``Robustness of helical magnetic structure under external pressure in Fe-doped MnNiGe alloy''}

\begin{figure}[h]
\centering
\includegraphics[width = 8 cm]{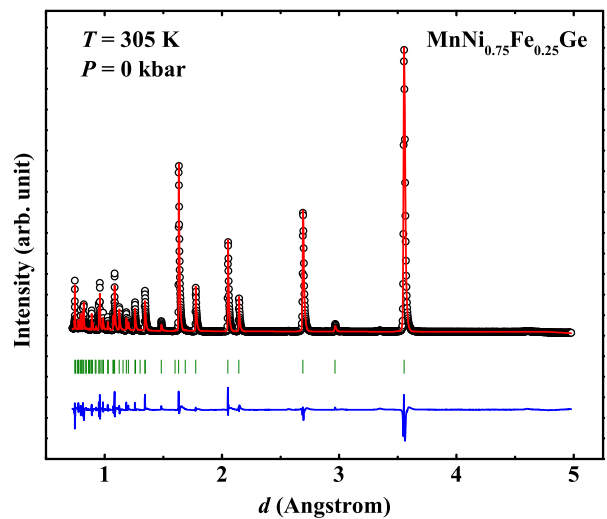}
\caption{(FIG. S1.) Neutron diffraction data recorded at 305 K in ambient condition are plotted along with the calculated pattern, difference pattern and the Bragg positions for the hexagonal nuclear phase.}
\label{npd}
\end{figure}

\begin{figure}[b]
\vskip 10mm
\centering
\includegraphics[width = 8 cm]{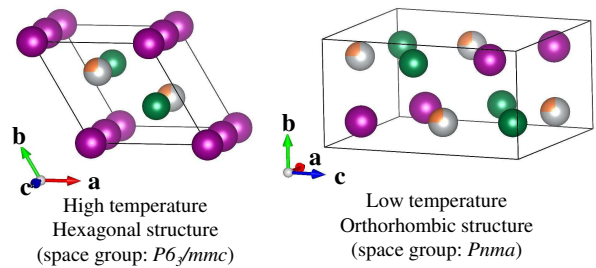}
\caption{(FIG. S2.) Illustration of the hexagonal and orthorhombic nuclear structure of MnNi$_{0.75}$Fe$_{0.25}$Ge alloy at 305 K and 1.5 K respectively. Here the purple, green, and white/orange color indicate Mn, Ge, and Ni/Fe atoms respectively.}
\label{str}
\end{figure}

\begin{table*}[t]
\centering
\caption{(TABLE S1.) Various structural parameters in ambient (at 1.5 K) and high pressure (at 10 K) conditions.}
\resizebox{\textwidth}{!}{
\begin{tabular}{cccccc|cccccc}
\hline
\hline
\multicolumn{5}{c}{Ambient pressure at $T$ = 1.5 K} & & \multicolumn{5}{c}{6 kbar pressure at $T$ = 10 K} &\\
\hline
\multicolumn{12}{c}{Orthorhombic structure with space group $Pnma$} \\
\hline
\multicolumn{2}{c}{$a$ = 5.9941(7) \AA} & \multicolumn{2}{c}{$b$ = 3.7497(5) \AA} & $c$ = 7.1236(5) \AA & & \multicolumn{2}{c}{$a$ = 5.9794(8) \AA} & \multicolumn{2}{c}{$b$ = 3.7430(7) \AA} & {$c$ = 7.1147(4) \AA} \\
\multicolumn{2}{c}{$R_p$ = 5.33\%} & \multicolumn{2}{c}{$R_{wp}$ = 4.01\%} & Phase fraction = 99.52\% & & \multicolumn{2}{c}{$R_p$ = 0.32\%} & \multicolumn{2}{c}{$R_{wp}$ = 0.16\%} & Phase fraction = 99.54\% &  \\
\hline
\ Type & Site & $x$ & $y$ & $z$ & $B_{\rm iso}$ & Type & Site & $x$ & $y$ & $z$ & $B_{\rm iso}$\\
\hline
 Mn & 4$c$ & 0.0300(4) & 0.2500(0) & 0.1805(2) & 1.6771(1) & Mn & 4$c$ & 0.1147(9) & 0.2500(0) & 0.2417(8) & 1.2532(1)\\
Ni    &      4$c$    &    0.1497(4) & 0.2500(0) & 0.5571(9) & 0.8244(2)   &    Ni   &     4$c$    &    0.1450(4) & 0.2500(0) & 0.5468(1) & 1.6301(2)\\
\ Fe     &     4$c$   &     0.1497(4) & 0.2500(0) & 0.5571(9) & 0.8244(2)   &    Fe    &     4$c$    &    0.1450(4) & 0.2500(0) & 0.5468(1) & 1.6301(2)\\
\ Ge    &     4$c$    &    0.7586(3) & 0.2500(0) & 0.6253(2) & 0.9425(2)   &    Ge    &     4$c$    &    0.8149(3) & 0.2500(0) & 0.6507(2) & 1.5124(1)\\
\hline
\hline
\multicolumn{12}{c}{Hexagonal structure with space group$P6_3/mmc$} \\
\hline
\multicolumn{2}{c}{$a$ = 4.0888(8)  \AA} & \multicolumn{2}{c}{$c$ = 5.3199(9)  \AA} & $\gamma$ = 120$^{\circ}$ & & \multicolumn{2}{c}{$a$ = 4.0671(9)  \AA} & \multicolumn{2}{c}{$c$ = 5.3430(6)  \AA} & $\gamma$ = 120$^{\circ}$ \\
\multicolumn{2}{c}{$R_p$ = 7.14\%} & \multicolumn{2}{c}{$R_{wp}$ = 8.16\%} & Phase fraction = 0.48\% & & \multicolumn{2}{c}{$R_p$ = 2.50\%} & \multicolumn{2}{c}{$R_{wp}$ = 1.32\%} & Phase fraction = 0.46\% &  \\
\hline
\ Type & Site & $x$ & $y$ & $z$ & $B_{\rm iso}$ & Type & Site & $x$ & $y$ & $z$ & $B_{\rm iso}$\\
\hline
\ Mn    &     2$a$   &     0.0000(0)  &  0.0000(0)  &   0.0000(0) &  1.3780(1)   &    Mn    &     2$a$    &    0.0000(0)  &  0.0000(0)  &   0.0000(0)  & 1.428(4)\\
\ Ni     &     2$d$   &     0.3333(0)   &   0.6666(0)    &   0.7500(0)  &   0.0720(3)  &     Ni    &      2$d$    &    0.3333(0) &     0.6666(0)   &    0.7500(0)  &   0.1720(4)\\
\ Fe    &      2$d$   &     0.3333(0)  &    0.6666(0)   &    0.7500(0)   &  1.0485(2)   &    Fe    &      2$d$   &     0.3333(0)  &    0.6666(0)    &   0.7500(0)  &   1.2480(3)\\
\ Ge    &     2$c$    &    0.3333(0)   &   0.6666(0)   &    0.2500(0)  &   0.6193(1)  &     Ge    &     2$c$    &    0.3333(0)  &    0.6666(0)   &    0.2500(0)   &  0.6994(5)\\
\hline
\hline
\end{tabular}
}
\end{table*}

\begin{figure*}[t]
\centering
\includegraphics[width = 12 cm]{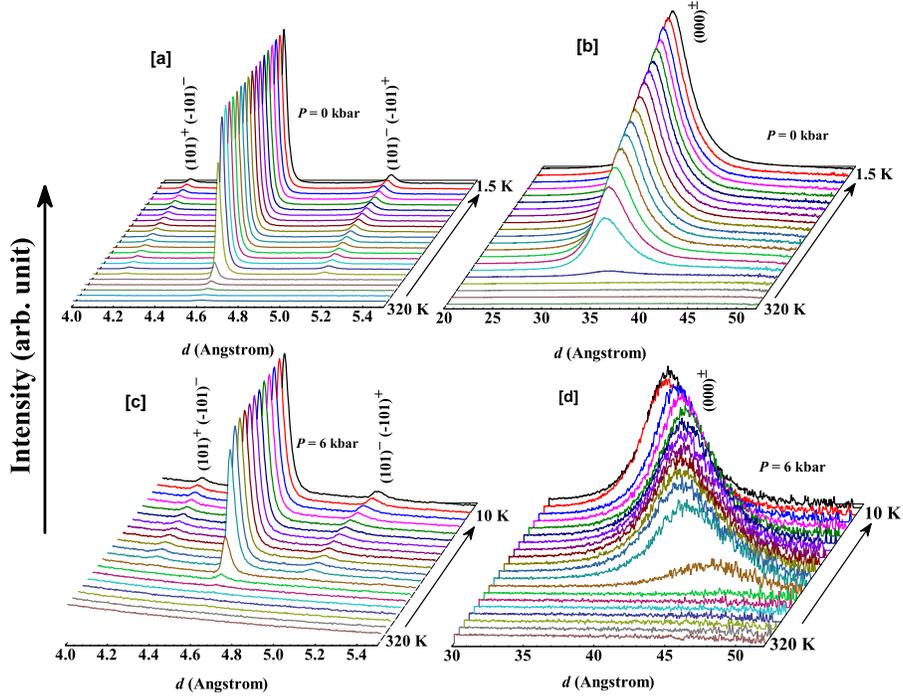}
\caption{(FIG. S3.) Restricted views of the temperature evolution of (101)$^+$, (-101)$^-$, (101)$^-$, (-101)$^+$ and (000)$^{\pm}$ magnetic satellite reflections in ambient (a and b) as well as in presence of 6 kbar of external pressure (c and d).}
\label{npd1}
\end{figure*}

\begin{figure*}[t]
\centering
\includegraphics[width = 14 cm]{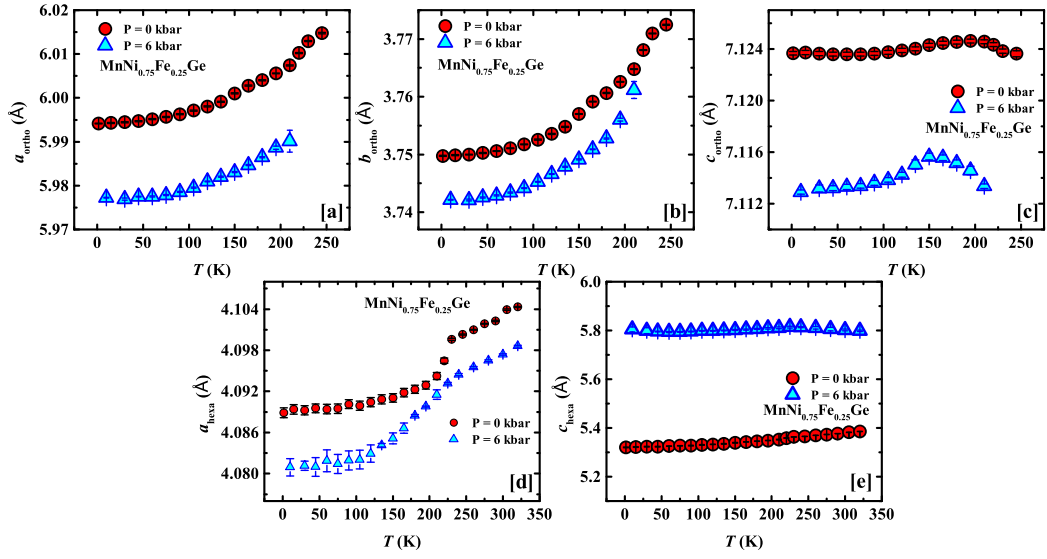}
\caption{(FIG. S4.) Temperature ($T$) variations of orthorhombic (a, b and c) and hexagonal (d and e) lattice parameters in ambient as well as in presence of 6 kbar of external pressure.}
\label{para}
\end{figure*}

\end{document}